\documentclass[12pt]{amsart}
\usepackage{amssymb}
\usepackage{txfonts}
\usepackage{amsfonts}
\usepackage{mathrsfs}
 \theoremstyle{definition}
 
 \theoremstyle{remark}

 \numberwithin{equation}{section}

\begin{document}
\title[THE CHSH-TYPE INEQUALITIES FOR INFINITE-DIMENSIONS]
{THE CHSH-TYPE INEQUALITIES FOR INFINITE-DIMENSIONAL QUANTUM SYSTEMS}

\author{Yu Guo}
\address{
Department of Mathematics, Shanxi Datong University, Datong, 037009, China}
 \email{guoyu3@aliyun.com; yuguophys@aliyun.com}



\begin{abstract}

By establishing CHSH operators and CHSH-type inequalities,
we show that any entangled pure state in
infinite-dimensional systems is entangled in a
$2\otimes2$ subspace.
We find that, for infinite-dimensional systems,
the corresponding properties are similar to that of the two-qubit case:
(i) The CHSH-type inequalities provide a sufficient and necessary
condition for separability of pure states;
(ii) The CHSH operators satisfy the Cirel'son
inequalities;
(iii) Any state which violates one of these Bell inequalities
is distillable.

\end{abstract}

\maketitle

\section{Introduction}

One of the entanglement problems is to decide whether or not a
prepared state is entangled \cite{MN,Horodecki1,Guhne,Zhao,Zhang,Wangyinzhu,HJ,HJ4,guojpa,GY}.
The first
condition of entanglement was given by Bell in 1964 \cite{Bell}.
Werner \cite{Werner} first pointed out that separable states must satisfy all
possible \emph{Bell inequalities}. One of the most important
Bell inequalities is the so-called \emph{CHSH inequalities} investigated by
Clauser, Horne, Shimony, and Holt \cite{Cla} in 1969.

The CHSH inequalities have been studied
intensively since they
provide a way of experimentally testing the local hidden variable
model as an independent hypothesis separated from the quantum
formalism for two-qubit systems. \if Cirel'son \cite{Cirelson} showed
in 1980 that all the expectation values of the CHCH operators are
not larger than $2\sqrt{2}$.\fi
In 1991, Gisin showed that the CHSH
inequalities are not only necessary but also sufficient for
separability of two-qubit pure states \cite{Gisin}. A short later,
Gisin and Peres found that this fact can be extended into
two-qudit systems \cite{Gisin2}. And for the three-qubit case, Chen
\emph{et al.} proved that all pure entangled states violate some Bell
inequality \cite{Chen}. Since the Bell inequalities just provide a
necessary condition for separability of mixed states, one might want
to know which states violate the Bell inequalities.
Consequently, it is verified that any two-qubit \cite{Horodeckir} or
three-qubit \cite{Lee} state violating a specific Bell inequality is
distillable. This result also holds for $N$-qubit systems whenever the state
is regarded as a bipartite state associated with the corresponding
bipartite decomposition of the system \cite{Acin1,Acin2}.

It is worth mentioning that the Bell inequalities can be used in
verifying the security of quantum key distribution protocols
\cite{Ek} and that distillation of quantum entanglement plays a key
role in quantum information processing \cite{MN}.
In addition, one may even apply the CHSH-type inequality to
one-body quantum system when investigating quantum contextuality \cite{Yusixia}.
Very recently, M. Li
\emph{et al}. \cite{Lim} proposed CHSH-type inequalities for any
finite-dimensional composite system and showed that (i) a pure sate is
entangled if and only if some of the CHSH-type inequalities are violated and (ii) violating the
CHSH-type inequalities implies that the state is distillable.
(Note here that Ref.~\cite{Lim} was originally devoted to
extend the Gisin's Theorem to higher dimensional multipartite case,
however, it did not complete the proof. Indeed it
proposed CHSH-type inequalities, see Ref.~\cite{Chenjingling} for detail.)
Inspired by this, in this paper we discuss these issues for infinite-dimensional systems.
(Infinite-dimensional quantum systems are always related to the continuous variable systems,
such as harmonic oscillator, which has infinite eigenstates (or Fock states).)

This paper is organized as follows.
In section 2 we construct the CHSH
operators for infinite-dimensional systems. By virtue of the
concurrence for infinite-dimensional systems \cite{Guo3}(i.e.,
the concurrence of a pure state is zero if and
only if it is separable), we can use an argument similar to that in
Ref.~\cite{Lim} to prove the generalized CHSH-type inequalities. In addition, we find
that the generalized CHSH operators also satisfy the Cirel'son inequalities.
In section 3,
a sufficient condition for distillation of entanglement in infinite-dimensional
systems is proposed by reducing the given state to the
two-qubit state and by the virtue of concurrence. Consequently, we conclude
that any entangled pure sate in infinite-dimensional system
is distillable. A final summary in the last section concludes this paper.

\section{The CHSH-type inequalities for infinite-dimensional composite systems}

Recall that, the CHSH operators for two-qubit quantum systems are
constructed via the Pauli matrices, that is
\begin{equation}
\mathcal{B}
=A_1\otimes B_1+A_1\otimes B_2+A_2\otimes B_1-A_2\otimes
B_2,
\end{equation}
 where $A_i=\vec{a_i}\cdot
\vec{\sigma_A}=a_i^x\sigma_A^x+a_i^y\sigma_A^y+a_i^z\sigma_A^z$,
$B_j=\vec{b_j}\cdot
\vec{\sigma_B}=b_j^x\sigma_B^x+b_j^y\sigma_B^y+b_j^z\sigma_B^z$;
$\vec{a_i}=(a_i^x,a_i^y,a_i^z)$ and
$\vec{b_j}=(b_j^x,b_j^y,b_j^z)$ are real vectors
satisfying $(a_i^x)^2+(a_i^y)^2+(a_i^z)^2=1$,
$(b_j^x)^2+(b_j^y)^2+(b_j^z)^2=1$; $\sigma_{A/B}^{x,y,z}$ are Pauli
matrices. The CHSH-type inequalities read as
\begin{equation}
|\langle\mathcal{B}\rangle|\leq2.
\end{equation}
Namely, if there exist local hidden variable models describing the
system, then the inequalities in (2.2) must hold.

In the following, we construct CHSH operators for
infinite-dimensional quantum systems.

\subsection{Bipartite systems}

Consider a system consisting of two parts, labeled A and B, with state
spaces $H_A$ and $H_B$ ($\dim H_A\otimes H_B=+\infty$) respectively. For the
fixed basis $\{|i\rangle\}_{i=1}^{+\infty}$ of $H_A$, let
\if false
$$\{L_{\alpha}^A\}
=\{|i\rangle\langle j|-|j\rangle\langle i|,\ 1\leq i<j\leq+\infty\}.$$
For example, we write
\fi
$L_1^A=|1\rangle\langle2|-|2\rangle\langle1|$,
$L_2^A=|1\rangle\langle3|-|3\rangle\langle1|$,
$L_3^A=|2\rangle\langle3|-|3\rangle\langle2|$,
$L_4^A=|1\rangle\langle4|-|4\rangle\langle1|$,
$L_5^A=|2\rangle\langle4|-|4\rangle\langle2|$,
$L_6^A=|3\rangle\langle4|-|4\rangle\langle3|$,
$L_7^A=|1\rangle\langle5|-|5\rangle\langle1|$, $\cdots$,
and $L_\alpha^A=|i\rangle\langle j|-|j\rangle\langle i|$
with $\alpha=(j-2)!+i$ whenever $j\geq5$. Similarly,
we can define $\{L_\beta^B\}$. The operators $L_\alpha^A$
(resp. $L_\beta^B$) have two none zero entries. We
define $A_i^\alpha$ (resp. $B_j^\beta$) from $L^A_\alpha$
(resp. $L^B_\beta$) by replacing the four entries on the
positions of the nonzero two rows and two columns of $L_\alpha^A$
(resp. $L_\beta^B$) with the corresponding four entries
of the matrix $\vec{a_i}\vec{\sigma_A}$
(resp. $\vec{b_j}\vec{\sigma_B}$), and
keeping the other entries of $A_i^\alpha$ (resp. $B_j^\beta$)
zero. We define the CHSH operators to be
\begin{equation}
\mathcal{B}_{\alpha\beta}
=\tilde{A}_1^\alpha\otimes \tilde{B}_1^\beta+
\tilde{A}_1^\alpha\otimes \tilde{B}_2^\beta+
\tilde{A}_2^\alpha\otimes \tilde{B}_1^\beta-
\tilde{A}_2^\alpha\otimes \tilde{B}_2^\beta,
\end{equation} where
$\tilde{A}_i^\alpha=L^A_\alpha A_i^\alpha (L^A_\alpha)^\dagger$,
$\tilde{B}_j^\beta=L^B_\beta B_j^\beta (L^B_\beta)^\dagger$, $i$, $j=1$, 2.

Let us extend this idea to the multipartite case in the following.

\subsection{Multipartite systems}

In this subsection, we discuss the multipartite system $H_1\otimes
H_2\otimes\cdots\otimes H_m$, $\dim H_1\otimes
H_2\otimes\cdots\otimes H_m=+\infty$. Any pure state
$|\psi\rangle\in H_1\otimes H_2\otimes\cdots\otimes H_m$ can be
represented by
\begin{equation}
|\psi\rangle
=\sum\limits_{i_1,i_2,\cdots,i_m}a_{i_1i_2\cdots i_m}|i_1i_2\cdots i_m\rangle.
\end{equation}

Let $\alpha$ and $\alpha'$ (respectively, $\beta$ and $\beta'$) be subsets
of the sub-indices of $a$, associated to the
same sub-vector spaces but with different summing indices. $\alpha$
(or $\alpha'$) and $\beta$ (or $\beta'$) span the whole space of the
given sub-index of $a$. Possible combinations of the indices of
$\alpha$ and $\beta$ can be equivalently understood as a kind of
bipartite decomposition of the $m$ subsystems, say part A$'$ and
part B$'$, containing $s$ and $t=m-s$ subsystems respectively.

For a fixed bipartite decomposition, the system can be regarded as a
bipartite system. Let $L_\alpha^A$ and $L_\beta^B$ be the operators
defined as above. Let $\rho=|\psi\rangle\langle\psi|$,
then
\begin{equation}
\rho_{\alpha\beta}^p
=\frac{L_\alpha^A\otimes L_\beta^B\rho(L_\alpha^A)^\dagger\otimes(L_\beta^B)^\dagger}
{\|L_\alpha^A\otimes L_\beta^B\rho(L_\alpha^A)^\dagger\otimes(L_\beta^B)^\dagger\|_{\rm Tr}}
\end{equation}
is a ``two-qubit'' pure state, where $p$ denotes the fixed bipartite decomposition.
Now, for any given bipartite decomposition, we can define the
corresponding CHSH operators
\begin{equation}\mathcal{B}_{\alpha\beta}^p
=\tilde{A}_1^\alpha\otimes \tilde{B}_1^\beta+
\tilde{A}_1^\alpha\otimes \tilde{B}_2^\beta+
\tilde{A}_2^\alpha\otimes \tilde{B}_1^\beta-
\tilde{A}_2^\alpha\otimes \tilde{B}_2^\beta,
\end{equation}
where
$\tilde{A}_i^\alpha=L^A_\alpha A_i^\alpha (L^A_\alpha)^\dagger$ and
$\tilde{B}_j^\beta=L^B_\beta B_j^\beta (L^B_\beta)^\dagger$
are the self-adjoint operators defined as in Eq.(2.3).

By the feature of concurrence, M. Li \emph{et al}. generalized the CHSH
operators to arbitrary finite-dimensional cases \cite{Lim}.
Very recently, we proposed concurrence for infinite-dimensional cases:

\vspace*{12pt}
\noindent{\bf Definition 2.1.}(Guo \emph{et al.} \cite{Guo3}) \ Let $|\psi\rangle\in H_A\otimes H_B$ be a pure state with $\dim
H_A\otimes H_B=+\infty$, and let $\{|i\rangle\}$ and $\{|j\rangle\}$ be the bases of
$H_A$ and $H_B$, respectively. Write
$|\psi\rangle=\sum\limits_{i,j}a_{ij}|i\rangle|j\rangle$. Then
\begin{equation}
C(|\psi\rangle)=\sqrt{2(1-{\rm Tr}(\rho_A^2))},\
\rho_A={\rm Tr}_B(|\psi\rangle\langle\psi|)
\end{equation}
 is called
the concurrence of $|\psi\rangle$. Moreover,
\begin{equation}
C(|\psi\rangle)
=\sqrt{\sum\limits_{i,j,k,l}|a_{ik}a_{jl}-a_{il}a_{jk}|^2}.
\end{equation}
\vspace*{12pt}

It is clear that $C(|\psi\rangle)=0$ if and only if $|\psi\rangle$
is separable.
Letting $\rho_{\alpha\beta}
=\frac{L_\alpha^A\otimes L_\beta^B\rho (L_\alpha^A)^\dag\otimes (L_\beta^B)^\dag}
{\|L_\alpha^A\otimes L_\beta^B\rho (L_\alpha^A)^\dag\otimes (L_\beta^B)^\dag\|_{\rm Tr}}$, one can check that
\begin{equation}
C(|\psi\rangle)=\sqrt{\sum_\alpha\sum_\beta[C(\rho_{\alpha\beta})]^2}.
\end{equation}

With this basic idea in mind, similar to that of the finite-dimensional case \cite{Fei2001},we define concurrence for
infinite-dimensional multipartite systems.

\vspace*{12pt}
\noindent{\bf Definition 2.2.} \ Let $\{|i_k\rangle\}$ be the basis of $H_k$,
$k=1$, 2, $\dots$, $m$, $|\psi\rangle=
\sum\limits_{i_1,i_2,\cdots,i_m}a_{i_1i_2\cdots i_m}|i_1i_2\cdots
i_m\rangle \in H_1\otimes H_2\otimes\cdots\otimes H_m$ be a pure state with $\dim
H_1\otimes H_2\otimes\cdots\otimes H_m=+\infty$. Then
\begin{equation}
C(|\psi\rangle)=\sqrt{\frac{1}{2^{m-1}-1}
\sum\limits_p\sum\limits_{\alpha,\alpha^\prime,\beta,\beta^\prime}
|a_{\alpha\beta}a_{\alpha^\prime\beta^\prime}
-a_{\alpha\beta^\prime}a_{\alpha^\prime\beta}|^2}
\end{equation}
 is
called the concurrence of $|\psi\rangle$. Here $\alpha$ and
$\alpha^\prime$ (respectively, $\beta$ and $\beta^\prime$) are subsets of the
sub-indices of $a$, associated to the same sub Hilbert space but
with different summing indices; $\alpha$ (or $\alpha^\prime$) and
$\beta$ (or $\beta^\prime$) span the whole space of a given subindex
of $a$; and where $\sum\limits_p$ stands for the summation over all
possible combinations of the indices of $\alpha$ and $\beta$.
\vspace*{12pt}

From Definition 2.2, one can obtain that $C(|\psi\rangle)=0$ if and
only if $|\psi\rangle$ is fully separable (namely, $|\psi\rangle
=|\psi_1\rangle\otimes|\psi_2\rangle\otimes\cdots\otimes|\psi_m\rangle$).
It is easy to see that
\begin{equation}
C(|\psi\rangle)=\sqrt{\frac{1}{2^{m-1}-1}
\sum\limits_p\sum\limits_{\alpha,\beta}
[C(\rho^p_{\alpha\beta})]^2}.
\end{equation}

Based on Eq.(2.6) and Eq.(2.8),  using an argument similar to that in Ref.~\cite{Lim} one can obtain the
following results.

\vspace*{12pt}
\noindent{\bf Theorem 2.3.} \  Let $\rho$ be a bipartite pure state acting
on $H_A\otimes H_B$ with $\dim H_A\otimes H_B=+\infty$. Then $\rho$ is
separable if and only if $\rho$ satisfies all Bell inequalities,
i.e.,
\begin{equation}
|\langle \mathcal{B}_{\alpha\beta}\rangle|\leq2
\end{equation}
for all CHSH operators as in Eq.(2.3).

\vspace*{12pt}
\noindent{\bf Theorem 2.4.} \ Let $\rho$ be a multipartite pure state
acting on $H_1\otimes H_2\otimes\cdots\otimes H_m$ with $\dim H_1\otimes
H_2\otimes\cdots\otimes H_m=+\infty$. Then $\rho$ is separable if
and only if $\rho$ satisfies all Bell inequalities, i.e.,
\begin{equation}
|\langle \mathcal{B}_{\alpha\beta}^p\rangle|\leq2
\end{equation}
for all CHSH operators as in Eq.(2.4).
\vspace*{12pt}

Let $\mathcal{S}_{s-p}$ be the
set of all separable pure states. Observe that, any separable state $\rho$ acting on
$H_A\otimes H_B$ admits a representation of the Bochner integral
\cite{Holevo}
\begin{equation}
\rho=\int_{\mathcal{S}_{s-p}}\varphi(\rho^A\otimes\rho^B)d\mu(\rho^A\otimes\rho^B),
\end{equation}
where $\mu$ is a Borel
probability measure on $\mathcal{S}_{s-p}$,
$\rho^A\otimes\rho^B\in\mathcal{S}_{s-p}$, and
$\varphi:\mathcal{S}_{s-p}\rightarrow\mathcal{S}_{s-p}$ is a
measurable function. (Note that the above definition is
equivalent to that given in \cite{Werner,HP2}.) Thus, for any CHSH operator
$\mathcal{B}_{\alpha\beta}$, it is straightforward that
$$\begin{array}{rl}|\langle {\mathcal B}_{\alpha\beta}\rangle|\leq&
\int_{\mathcal{S}_{s-p}}|{\rm Tr}[\varphi(\rho^A\otimes\rho^B)
{\mathcal B}_{\alpha\beta}]|d\mu(\rho^A\otimes\rho^B)\\
\leq&2.\end{array}$$
That is, all separable mixed states also satisfy the
inequalities in (2.9). Similarly, all fully separable multipartite mixed states
satisfying the inequalities in (2.10).

Going further, we discuss the relation between the CHSH operators
and \emph{entanglement witnesses}. It is shown in  \cite{HM1}
that a given state is entangled if and only if there exists
at least one entanglement witness detecting it. An entanglement
witness $W$ is a self-adjoint operator (also sometimes called a Hermitian
operator) acting on $H_A\otimes H_B$ that satisfies ${\rm
Tr}(W\sigma)\geq0$ for all separable sates $\sigma$ and ${\rm
Tr}(W\rho)<0$ for at least one entangled state $\rho$. Now, letting
\begin{equation}
W_{\alpha\beta}=2I-\mathcal{B}_{\alpha\beta},
\end{equation}
it is clear that ${\rm Tr}(W\sigma)\geq0$ for all separable
states and ${\rm Tr}(W\rho)<0$ if and only if
$|\langle \mathcal{B}_{\alpha\beta}\rangle|\leq 2$.
Thus, $W_{\alpha\beta}$ is an entanglement witness provided that
$2I-\mathcal{B}_{\alpha\beta}$ is not positive. Form this point of view,
the CHSH operators are entanglement witnesses that
can detect all entangled pure states.

For two-qubit systems, the CHSH operators satisfy
the Cirel'son inequality \cite{Cirelson}
\begin{equation}
|\langle\mathcal{B}\rangle|\leq2\sqrt{2}
\end{equation}
for all states $\rho$ and all CHSH operators $\mathcal{B}$. The
state $\rho=\frac{1}{2}(|01\rangle-|10\rangle)(\langle01|-\langle10|)$
saturates the Cirel'son bound, i.e., $|{\rm Tr}(\mathcal{B}\rho)|=2\sqrt{2}$.

\vspace*{12pt}
\noindent{\bf Theorem 2.5.} \ The Cirel'son inequality holds for composite systems of any dimension.

\begin{proof}
Now we consider the generalized CHSH operators (for both finite- and
infinite-dimensional cases).
We write $P_\alpha=|i\rangle
\langle i|+|j\rangle\langle j|$ whenever $L_{\alpha}=|i\rangle\langle j|-
|j\rangle\langle i|$ and define $P_\beta$ similarly. Then we consider
the squares of $\tilde{A}_i^\alpha$ and $\tilde{B}_j^\beta$:
$(\tilde{A_i^{\alpha}})^2
=L_{\alpha} A_i^{\alpha} L_{\alpha}^{\dag} L_{\alpha} A_i^{\alpha} L_{\alpha}^{\dag}
=L_{\alpha} (A_i^{\alpha})^2 L_{\alpha}^{\dag}=P_{\alpha}$,
$(\tilde{B_j}^{\beta})^2=P_\beta$. Using the fact
$$\begin{array}{rl}&\tilde{A}_1^\alpha\otimes \tilde{B}_1^\beta+
\tilde{A}_1^\alpha\otimes \tilde{B}_2^\beta+
\tilde{A}_2^\alpha\otimes \tilde{B}_1^\beta-
\tilde{A}_2^\alpha\otimes \tilde{B}_2^\beta\\
=&\frac{1}{\sqrt{2}}((\tilde{A}_1^\alpha)^2\otimes P_\beta
+(\tilde{A}_2^\alpha)^2\otimes P_\beta\\
&+P_\alpha\otimes(\tilde{B}_1^\beta)^2
+P_\alpha\otimes(\tilde{B}_2^\beta)^2)\\
&-\frac{\sqrt{2}-1}{8}((\sqrt{2}+1)
(\tilde{A}_1^\alpha\otimes P_\beta-P_\alpha\otimes\tilde{B}_1^\beta)\\
&+\tilde{A}_2^\alpha\otimes P_\beta-P_\alpha\otimes\tilde{B}_2^\beta)^2\\
&-\frac{\sqrt{2}-1}{8}((\sqrt{2}+1)
(\tilde{A}_1^\alpha\otimes P_\beta-P_\alpha\otimes\tilde{B}_2^\beta)\\
&-\tilde{A}_2^\alpha\otimes P_\beta-P_\alpha\otimes\tilde{B}_1^\beta)^2\\
&-\frac{\sqrt{2}-1}{8}((\sqrt{2}+1)
(\tilde{A}_2^\alpha\otimes P_\beta-P_\alpha\otimes\tilde{B}_1^\beta)\\
&+\tilde{A}_1^\alpha\otimes P_\beta+P_\alpha\otimes\tilde{B}_2^\beta)^2\\
&-\frac{\sqrt{2}-1}{8}((\sqrt{2}+1)
(\tilde{A}_2^\alpha\otimes P_\beta+P_\alpha\otimes\tilde{B}_1^\beta)\\
&-\tilde{A}_1^\alpha\otimes P_\beta-P_\alpha\otimes\tilde{B}_1^\beta)^2\\
\leq&\frac{1}{\sqrt{2}}((\tilde{A}_1^\alpha)^2\otimes P_\beta)
+(\tilde{A}_2^\alpha)^2\otimes P_\beta\\
&+P_\alpha\otimes(\tilde{B}_1^\beta)^2
+P_\alpha\otimes(\tilde{B}_2^\beta)^2)\\
=&2\sqrt{2}P_\alpha\otimes P_\beta\end{array}$$ we have
\begin{equation}
|\langle\mathcal{B}_{\alpha\beta}\rangle|\leq2\sqrt{2}
\end{equation}
since $|\langle P_\alpha\otimes P_\beta\rangle|\leq 1$.
Similarly, we have
\begin{equation*}
|\langle\mathcal{B}^p_{\alpha\beta}\rangle|\leq2\sqrt{2}.
\end{equation*}
\end{proof}

\section{CHSH-type inequalities and distillation}

For the finite-dimensional case, it is shown in Ref.~\cite{Lim}
that quantum states that violate the CHSH-type
inequalities must be distillable.
In this section, we discuss the infinite-dimensional case.

The famous singlet states play a crucial role in quantum information
theory since for many tasks, like teleportation and cryptography, one
ideally needs maximally entangled two-qubit states. However, in
laboratories we usually have mixed states due to imperfection of
operations and decoherence. It makes sense to transform the mixed
states to the useful singlet states. Recall that, \if in the
finite-dimensional bipartite system,\fi a state $\rho$ is defined to
be distillable if and only if the singlet state (e.g.
$|\psi^\pm\rangle=\frac{1}{\sqrt{2}}(|00\rangle\pm|11\rangle)$) can be
obtained from a finite number of copies of $\rho$ by LOCC.
In Ref.~\cite{Lim}, in the
finite-dimensional multipartite system, a multipartite state is
called distillable if and only if there exists some bipartite
decomposition of the system such that the state is distillable
whenever it is regarded as a bipartite state.

Any LOCC for an infinite-dimensional bipartite system admits the form of
\begin{equation}
\Lambda(\rho)
=\sum\limits_{i=1}^NA_i\otimes B_i\rho A_i^\dagger
\otimes B_i^\dagger
\end{equation}
where $A_i$ and $B_i$ are operators acting on $H_A$ and $H_B$ respectively,
$\sum\limits_{i=1}^NA_i^\dagger A_i\otimes B_i^\dagger B_i
=I_A\otimes I_B$, and
 the series converges in the strong operator
topology \cite{HJ} and $N$ may be $+\infty$.

For an infinite-dimensional multipartite system, we can define
distillation of entanglement as follows.

\vspace*{12pt}
\noindent{\bf Definition 3.1.} \  Let $H_i$, $i=1$, 2, $\dots$, $m$,
be complex separable Hilbert spaces with $\dim H_1\otimes
H_2\otimes\cdots\otimes H_m=+\infty$. A state $\rho$ acting on
$H_1\otimes H_2\otimes\cdots\otimes H_m$ is said to be distillable
if and only if there exists some bipartite decomposition of the
system such that $\rho$ is distillable whenever it is regarded as a
bipartite state $\check{\rho}=\rho^{A'B'}$ (acting on $H_{A'}\otimes
H_{B'}=H_{1'}\otimes H_{2'}\otimes\cdots\otimes H_{m'}$, where
$\{1',2',\dots,m'\}=\{1,2,\dots,m\}$), i.e., $\rho$ is distillable
if and only if there exists some LOCC $\Lambda$ and a finite number
$n$ such that $$\Lambda(\check{\rho}^{\otimes n})\ \mbox{\rm is a
singlet state.}$$
\vspace*{12pt}

It is obvious that a bipartite state $\rho$ is distillable if there
exist some projectors $P$ and $Q$ that map $H_A$ and $H_B$ into
${\mathbb C}^2$ and ${\mathbb C}^2$ respectively such that $P\otimes
Q\rho P\otimes Q$ is entangled.

The main result of this section is the following.

\vspace*{12pt}
\noindent{\bf Theorem 3.2.} \   Let $H_A$ and $H_B$ be complex separable Hilbert spaces with
$\dim H_A\otimes H_B=+\infty$. If the
state $\rho$ acting on $H_A\otimes H_B$ violates one of the
CHSH-type inequalities in (2.9), then $\rho$ is distillable. Moreover,
assume that $\dim H_1\otimes H_2\otimes\cdots\otimes H_m=+\infty$,
if the state $\rho$ acting on $H_1\otimes H_2\otimes\cdots\otimes
H_m$ violate one of the CHSH-type inequalities in (2.10), then $\rho$ is
distillable.

\begin{proof}
 We only need to prove the first part; the
multipartite case can be checked similarly.

If ${\rm Tr}(B_{\alpha_0\beta_0}\rho)>2$ for some $\alpha_0$ and
$\beta_0$, then
$$\rho_{\alpha_0\beta_0}
=\frac{L_{\alpha_0}^A\otimes L_{\beta_0}^B\rho(L_{\alpha_0}^A)^\dagger
\otimes (L_{\beta_0}^B)^\dagger}
{\|L_{\alpha_0}^A\otimes L_{\beta_0}^B\rho(L_{\alpha_0}^A)^\dagger
\otimes (L_{\beta_0}^B)^\dagger\|_{\rm Tr}}$$ and $C(\rho_{\alpha_0\beta_0})>0$.
Write $L_{\alpha_0}=|i\rangle\langle j|-|j\rangle\langle i|$,
$L_{\beta_0}=|k\rangle\langle l|-|l\rangle\langle k|$. Let $P=AL_{\alpha_0}$
and $Q=BL_{\beta_0}$ with $A=|0_A\rangle\langle i|+|1_A\rangle\langle j|$,
$B=|0_B\rangle\langle k|+|1_B\rangle\langle l|$,
where $|0_{A,B}\rangle$ and $|1_{A,B}\rangle$
are orthonormal bases of $\mathbb{C}^2$. It is clear
that $P\otimes Q\rho P\otimes Q$ is a two-qubit state and
$C(P\otimes Q\rho P\otimes Q)=C(\rho_{\alpha_0\beta_0})$.
Therefore $\rho$ is distillable since any
two-qubit state is distillable.
\end{proof}

Like the finite-dimensional case, the infinite-dimensional PPT state
should never violate the CHSH-type inequalities in (2.9) or (2.10) since PPT
states are not distillable. In the other words, the CHSH operators
are entanglement witnesses that cannot detect any PPT entangled state.
In Ref.~\cite{Bennett}, it is shown that any entangled pure state in a
finite-dimensional system is distillable. \if Remak that, a pure state
is separable if and only if it is a PPT state \cite{GY}, thus,\fi
Together with Theorem 3.2,
we can conclude that:

\vspace*{12pt}

\noindent{\bf Proposition 3.3.}\  Let $H_A$ and $H_B$ be complex
separable Hilbert spaces with $\dim H_A\otimes H_B\leq+\infty$, and let
$|\psi\rangle\in H_A\otimes H_B$ be a pure state. Then $|\psi\rangle$
is distillable if and only if it is entangled.
\vspace*{12pt}

Namely, there are no bound entangled pure states in either finite- or
infinite-dimensional systems.

\section{Conclusion}

In summary,
the CHSH-type inequalities for
the infinite-dimensional systems were proposed. We showed that,
in nature, for any entangled pure state, there exists at least
one $2\otimes 2$ subspace such that the
associated state is entangled and therefore
the corresponding properties are valid.
(Note that, for mixed states, Sperling and Vogel showed in
Ref.~\cite{Sperling} that any bipartite entanglement in
infinite-dimensional systems can
be identified in finite-dimensional Hilbert spaces, even
though in general the reduction in the Hilbert space to finite
dimensions may lead to nonclassical artifacts.)
The generalized CHSH-type inequalities may help in the
measurable determination of quantum entanglement
experimentally since there is a one-to-one correspondence between
the CHSH operators and the entanglement witnesses.

{\bf Acknowledgments} The author wishes to thank the referees for their useful comments and
suggestions.
This work is supported by
the China
Postdoctoral Science Foundation funded project (2012M520603),
the National Natural Science Foundation of China
(11171249,  11101250),
the Natural Science Foundation of Shanxi
(2013021001-1, 2012011001-2),
the Research start-up fund for Doctors of Shanxi Datong University
(2011-B-01) and the Scientific and Technological
Innovation Programs of Higher Education Institutions in Shanxi
(20121015).

\end{document}